\documentclass[runningheads]{llncs}

\usepackage[T1]{fontenc}
\usepackage{graphicx}
\usepackage{listings}

\begin{document}

\title{User Experiences with MPI RMA and ULFM in a Resilient Key-Value Store Implementation}
\author{Claudia Fohry \and Rainer Fink}

\authorrunning{C. Fohry and R. Fink}
\institute{University of Kassel, Germany\\
\email{\{fohry|rainer.fink@student\}@uni-kassel.de}}

\maketitle 
\begin{abstract}
 As hardware failures such as node losses become increasingly common,
MPI programmers may want to save vulnerable data in a resilient
store. While third-party storage solutions such as Redis or the
Hazelcast IMap exist, a tailored, MPI-based store may be easier to
integrate and can be optimized for particular application needs.

This paper considers the implementation of such a store, which is
intended as a component in a resilient task-based runtime system
written in MPI. The store holds redundant data copies as key-value
pairs in the main memories of multiple processes. Since store access
operations, such as reads and writes, are naturally one-sided, we
implemented the store with passive target MPI RMA functions. Process aborts
are detected with the user-level failure mitigation (ULFM) extension
of Open MPI. After failures, the program recovers on the surviving
processes and continues with the intact data copies.

Our implementation proved difficult, since several proposed ULFM
functionalities for RMA have not yet been implemented. Even assuming
their existence, we think that the programming task could be
simplified. This paper describes our experiences, lists
functionalities that we missed, and explains a workaround that we
adopted in our implementation.


\end{abstract}

\section{Introduction}\label{sect:intro}

Due to the increasing size of supercomputers, hardware failures such
as node losses are becoming increasingly common. In MPI programs,
these failures ma\-nifest as process losses, which normally lead to
program abort.  While checkpointing libraries such as
DMTCP~\cite{dmtcp} and VeloC~\cite{veloc} can be used to save all or
relevant program data and to later restart the program from a saved
state, it is preferable to recover the running program without 
interrupting it.

Task-based parallel programming is well-amenable to this
approach since the data at
clearly defined task interfaces can be transparently saved by the
runtime system and, after a failure, be used for a localized recovery
at the surviving processes~\cite{unserFGCS,miaNestedCP,miaJohnWamta}. The approach, called task-level
checkpointing, has low overheads~\cite{unserSupervisionVgl}.

Task-level checkpointing has previously been implemented in task-based systems atop Resilient
X10~\cite{resiX10,mcoAlt} and the related APGAS
library for Java~\cite{unserSupervisionVgl,miaNestedCP,apgas}. These
systems explicitly support fault tolerance by raising an
exception after the occurence of a failure at a communication partner. Moreover, they
support the registration of error handlers at any desired processes, which are
guaranteed to be eventually invoked after failure. However, most task-based
systems are implemented with C/C++ and MPI, where such
facilities are not available.

An important aspect in the realization of task-level checkpointing is
the actual storage of checkpoints and other vulnerable program
data. In task-based systems that employ work stealing, these data are
written frequently. The mentioned implementations therefore use an
in-memory store, the Hazelcast IMap~\cite{hazelcast}, which is a
resilient, distributed, key-value store with support for
transactions. The IMap can be easily accessed from X10 and APGAS
programs, and can also be integrated into C++/MPI programs.

While the IMap and related systems such as Redis~\cite{redis} and
etcd~\cite{etcd} are efficient in general, they did not scale well in
larger-scale experiments with the above type of applications
(\cite{ruedigerStore,miaNestedCP}), especially when we used them from
MPI programs. A major drawback of the third-party stores is their
generality, due to which they cannot exploit application specifics. In
our case, such specifics include a focus on writes as opposed to
reads, and the binding of keys to a single owner.
Moreover, the interface between the MPI program and the third-party
store causes overheads.

The goal of this work has been the development of a resilient
key-value store that is written in MPI and optimized for particular
application needs. The development was not entirely successful,
however, since we encountered various obstacles along the way. Most of
these were due to the current status of the design and development of the User Level
Failure Mitigation~(ULFM~\cite{ulfm}) extension of MPI, which we
needed for failure detection and recovery. ULFM is provided by Open
MPI~\cite{openMPI}, but, unfortunately, the ULFM support for one-sided MPI
communication (OSC) is still documented as untested. Still,
OSC is the natural communication mechanism for a store.

The paper makes the following contributions:
\begin{itemize}
\item We outline two design options for a resilient store and describe
  required programming functionalities for their implementation.
  \item We discuss the realization of these functionalities with the
    specified ULFM functionalities for OSC  and point to some
    possible enhancements.
  \item We experimentally evaluate the current status of support for
    the needed functions in Open MPI, confirming that it is still
    insufficient.
  \item We describe a workaround that we adopted to implement our
    store and provide some preliminary, but promising experimental
    results on its performance.
\end{itemize}

The paper is structured as follows. Section~\ref{sect:reqs}
provides background on our application and works out specifics that can
be used for store optimization. Then, Section~\ref{sect:background}
summarizes OSC and ULFM.  Section~\ref{sect:design} outlines design
options for the store with focus on required programming
functionalities, discusses their support by OSC/ULFM, and assembles
possible enhancements. Thereafter, the current implementation status
is assessed in Section~\ref{sect:experiences}. In
Section~\ref{sect:workaround}, we describe our workaround, sketch our
implementation, and provide experimental results. The
paper finishes with related work and conclusions, in
Sections~\ref{sect:related} and~\ref{sect.conclusions}, respectively.

\section{Application and Requirements to Resilient Store}\label{sect:reqs}

While MPI-based resilient stores have various applications,
this paper focuses on the specific use case of task-level
checkpointing, which protects task-based parallel programs against
fail-stop failures of processes. We refer to a particular technique that
was originally developed for dynamic independent tasks (DIT)
under cooperative work stealing~\cite{ijnc,unserFGCS} and later extended to
nested fork-join programs (NFJ)~\cite{miaNestedCP} and to future-based
cooperation (FBC) under coordinated work
stealing~\cite{miaJohnWamta}. Below, we sketch the technique and work
out its specific properties regarding usage of the resilient store.

The technique assumes that tasks are executed by independent worker
processes. Each worker maintains a task queue, into which it inserts
child tasks and from where it takes tasks for execution.  Moreover, 
idle workers can steal tasks from the queue. Workers write checkpoints
independently, at regular time intervals, as well as
during work stealing at the victim and thief sides. The checkpoint
contents differ between the above settings, with typical contents being
the current task pool, open task frames awaiting results,
futures, and identities of victims and thieves. Any new
checkpoint replaces the previous one.  In addition to the checkpoints,
the loot is temporarily saved while a steal is in progress.

In all settings, only a few, setting-specific types of data are
written (e.g. checkpoints and loot). These types exist once per worker
(called the \emph{owner}) and are overwritten by subsequent
accesses. Thus, in any instantiation of our key-value store, a fixed
set of keys, encoding owner id and type, is sufficient.  Moreover, it
can be planned statically, on which process which owner's data are saved. The same holds for replica of store entries, which provide
redundancy and thus enable recovery after multiple failures.

While the keys are known statically, the values and their sizes are
not, since, for instance, the checkpoint size depends on the number of
tasks in the task queue.

Outside failures, the store entries are only written to (e.g. when a
checkpoint is saved), and deleted (e.g. loot after stealing), but they are
not read. Thus, the store design should focus on writes as opposed to
reads.

In the DIT and NFJ variants, each store entry is updated by its owner
only. In the FBC variant, victim data may be written by the thief, but
even in this setting it is guaranteed (by outer locking mechanisms)
that only a single worker at a time may update any particular store
entry. Thus, outside failures, there are no concurrent accesses to the
same store entry.

After failures, a shrinking localized recovery is performed. Only a
few workers participate in this recovery, whereas the majority of
workers just continues task processing. More specifically, a
predetermined buddy worker and recent victims/thieves access data owned by the failed
worker. Additionally, if a second failure occurs, the
buddy may be replaced by another buddy. These workers must
coordinate their activities, and also take into account that store
access operations from the failed worker may arrive late. Therefore,
reference~\cite{ijnc} introduced a quite complex recovery protocol. It
includes transactions, which combine reads, writes, and conditionals.

During recovery, lost data and replicas are replaced and redistributed,
to be prepared for future failures. After recovery, the program
continues on the surviving processes (shrinking recovery).

\section{Background}\label{sect:background}

The Message Passing Interface (MPI) defines a comprehensive set of
functions for interprocess communication. MPI was
originally designed for single-threaded processes, but newer versions
support the simultaneous invocation of MPI functions by different
threads of a process. All MPI communication is accomplished in
\emph{communicators} (type \verb|MPI_Comm|), which are distributed
data structures that represent groups of processes and hold
reachability information.  Within communicators, \emph{collective
operations} such as broadcast can be executed. For these, \emph{all}
processes of the communicator must call the same function.  Most MPI
communication is pairwise and \emph{two-sided}, which means that both a sender
and a receiver process must invoke matching functions such as
\verb|MPI_Send| and \verb|MPI_Recv|.  In \emph{blocking} MPI communication, a
function call returns only when the function has (essentially)
finished, whereas in \emph{nonblocking} communication, the call
returns immediately and the actual work is done in the background.

\paragraph{Remote Memory Access (RMA)}: 
Since about the mid-nineties, MPI has supported RMA with \emph{one-sided
communication} (OSC) functions. In OSC, only one process calls a communication function and provides
all parameters.  OSC resembles shared-memory
programming, and is thus a natural choice for our resilient
store. Correspondlingly, this paper makes heavy use of MPI OSC.

As a prerequisite to OSC, all processes of a communicator must
collectively create a \emph{window} (type \verb|MPI_Win|) by calling a
function such as \verb|MPI_Win_create|. All window creation functions
are blocking.  Each participating process contributes a memory block to the call,
and the window is then composed of these individual memories. The
individual memory blocks can differ in size, and each process can
later change the size of its local block by calling
\verb|MPI_Win_attach|. Moreover, each process can access its local
block via C/C++ functions, whereas remote processes must use MPI
window access functions.

Window access functions have an argument of type \verb|MPI_Win|, but
they do not have an argument of type \verb|MPI_Comm|.  Basic \emph{communication
functions} include \verb|MPI_Put| and \verb|MPI_Get| for writing and
reading data, respectively. Moreover, atomic operations can be
realized, for instance with \verb|MPI_Compare_and_swap|.

All communication functions are nonblocking, and so the additional use
of \emph{synchronization functions} is necessary. These functions
define access epochs. Multiple communication functions
can be called in each epoch, and are guaranteed to be completed when the epoch ends.
Inside an epoch, function \verb|MPI_Win_flush| waits for the immediate completion
of all preceding communication.

MPI distinguishes active and passive target communication.  This paper
refers to passive target communication, since it gets closer to the
shared memory view of a store. In passive target
communication, an epoch at a particular window block is opened, for instance,
by a call to \verb|MPI_lock|, and closed by a call to
\verb|MPI_Unlock|. Locks can be exclusive (\verb|MPI_LOCK_EXCLUSIVE|)
or shared (\verb|MPI_LOCK_SHARED|), supporting the locking by a single
writer, or by multiple readers, respectively. For shared locks,
\verb|MPI_Win_lock_all| allows to lock all blocks of a window at
once. To enable optimizations, users may specify
assertions. Notably, with \verb|MPI_MODE_NOCHECK|, a user ensures that
there are no conflicting locks. This assertion can be used when mutual
exclusion is achieved by other means~\cite{mpi}.

\paragraph{User Level Failure Mitigation (ULFM):} All MPI communication functions require an argument of type \verb|MPI_Comm| or \verb|MPI_Win|, and all MPI communicators and windows have an associated error handler. By default, these handlers are set to \verb|MPI_ERRORS_ARE FATAL|, which causes any erroneous function call to abort the program. The individual handlers can be changed to \verb|MPI_ERRORS_RETURN|, and then erroneous function calls return an error code.

Usage of error codes for the handling of process aborts requires
additional facilities for the detection of aborts and for program
recovery. Such facilities are not yet part of  MPI, but a
corresponding extension, called ULFM, has been proposed~\cite{ulfm}. It is provided by the Open MPI implementation of MPI~\cite{openMPI}, to which we refer in the rest of this
paper. Fault-tolerant MPI programs must be started with
\begin{verbatim}
      mpirun ... --with-ft ulfm ...
\end{verbatim}

In ULFM, only affected processes are informed about an error. Most
importantly, communication functions such as \verb|MPI_Send| return
\verb|MPI_ERR_PROC_FAILED| when the communication partner has
failed. Furthermore, specific functionalities are available for
receive operations from unspecified senders (\verb|MPI_ANY_SOURCE|),
but these are not relevant for this paper.  If the receiver of an
error code wants to propagate the failure to other processes, it may
invalidate the corresponding communicator by calling
\verb|MPI_Comm_revoke|. Then, all processes of the communicator
receive error code \verb|MPI_ERR_REVOKED| when using the communicator.

Such failure propagation is necessary, for instance, before collective
operations can be executed again after a process abort. The actual
repair can be accomplished with function \verb|MPI_Comm_shrink|, which
must be called collectively.  Both blocking and nonblocking variants
exist. \verb|MPI_Comm_shrink| builds a new communicator from the
surviving processes and renumbers them. Its call should be combined
with \verb|MPI_Comm_agree| to ensure consistency.

While the above functionalities are meant for two-sided communication,
similar functionalities have been specified for OSC~\cite{ulfm}. They
include the return of \verb|MPI_ERR_PROC_FAILED| from window
synchronization functions such as \verb|MPI_Win_flush| and
\verb|MPI_Unlock|, and function \verb|MPI_Win_revoke|. Return codes
from communication functions such as \verb|MPI_Put| are
optional. Unfortunately, the ULFM functions for OSC have not yet been
fully implemented. When such support is requested, Open MPI issues the
warning
{\small
\begin{verbatim}
The selected 'osc' module 'rdma' is not tested for post-failure
operation, yet you have requested support for fault tolerance.
When using this component, normal failure free operation is expected;
However, failures may cause the application to abort, crash or deadlock.
\end{verbatim}
}
 
\section{Resilient Store Design Options and Implementation Issues}\label{sect:design}

In the following, we explain two design variants for our store: a
global recovery variant that has been implemented with minor
differences~\cite{maRainer}, and a localized recovery variant that
may improve on it, but has not yet been implemented.  For both,
we discuss required parallel programming functionalities and their
availability in the MPI and ULFM specifications.

In both variants, the MPI application creates the
store through a collective call of a store initialization
function. Arguments include the number of keys per worker, and the
initial size of the value per key. Thereafter, each
process can call store functions such as
\verb|store_put(key, value)| or \verb|store_delete(key)|. These functions are
blocking. Operation \verb|store_put| returns after the
value has been written to the master copy as well as to all replicas
in the resilient store. This synchronous design is simpler than the
more efficient alternative of an asynchronous design, in which the
function would return as soon as the master copy had been updated.
Internally, all store access operations use passive target MPI RMA
functions, and thus they do not interrupt the
processes at the targets.

\subsection{Resilient Store with Global Recovery}

From Section~\ref{sect:reqs}, the application enables a distinction between two operating modes of the store:
\begin{itemize}
\item \emph{Normal mode:} Outside failures, the store is accessed by
  write and delete operations only, and these are never
  invoked concurrently on the same entry.
  \item \emph{Transaction mode:} During recovery, the application invokes
    write, delete, and read operations, as well as
    transactions thereof, and these operations may be invoked
    concurrently by multiple processes.
\end{itemize}
Since normal mode takes up most of the time, we take the view that any possible
performance optimizations should be applied in this mode, no matter what their
price (in terms of performance) in transaction mode is. This led us to
the following design:

\paragraph{Normal mode:}  Each process writes its key-value pairs to a fixed set of other processes that are evenly spread across the machine. One of these processes holds the master copy, and the others hold replicas. The spreading follows a predetermined pattern, such that non-owners can localize the data as well. Since the number of keys per worker is fixed, one global window per key/replica is created with a collective operation of all processes. This window has datatype \verb|MPI_Byte|, and holds in its first entry the size of the currently stored value (0 if empty).

Operation \verb|store_put| (and analogously \verb|store_delete| )
internally performs \verb|MPI_Put| calls. For simplicity, the writing
process successively invokes this function for the master copy and for
all replicas. After each \verb|MPI_Put|, it calls
\verb|MPI_Win_flush| to ensure the correct ordering of the \verb|MPI_Put|s
from successive \verb|store_put| operations. According
to~\cite{27vonRainer}, \verb|MPI_Win_flush| is an efficient way to
synchronize the window accesses. Still, a blocking variant of
\verb|MPI_Put| would be more convenient for programmers.

Although, from an application correctness point of view, the above
flushes are sufficient, the MPI specification
requires us to define an epoch, which feels like programming ballast
that can be easily forgotten.  It may be specified with
\verb|MPI_Win_lock_all(MPI_MODE_NOCHECK, ...)| and
\verb|MPI_Win_unlock_all|. Currently, we use
\verb|MPI_Win_lock(MPI_LOCK_EXCLUSIVE,...)|,
assuming a single writer per entry.

A problem occurs when the data to be written are larger than the
window memory block. As noted in Section~\ref{sect:background}, the
block can be enlarged by a call to \verb|MPI_Win_attach|. However,
this function must be called by the target process of the
\verb|MPI_Put|, whereas the lack of memory is recognized by the
source. Unfortunately, in passive target OSC, there is no way for the
source to notify the target. While such notification could be
accomplished with nonblocking two-sided communication, posting
\verb|Irecv| at each potential target and occasionally testing for
incoming enlargement requests would significantly complicate the
program structure. A signalling mechanism between processes would be
desirable.

\paragraph{Transaction mode:}

When a failure occurs, the ULFM specification requires that
\verb|MPI_Win_flush| must return an error code. This code is received by any calling
process~R, e.g. during the write of a replica. The recovery, in
contrast, must in our application  be performed by some specific set of processes (buddy etc., see
Sect.~\ref{sect:background}) . These processes are not guaranteed to
communicate with the failed one, and so~R needs to inform them about the
failure. As noted above,~R can not signal them (apart from possibly not
knowing their identities),
and thus we decided to use ULFM's major failure
propagation mechanism, namely revoke. Only \verb|MPI_Win_revoke|, but not \verb|MPI_Comm_revoke|
is helpful here, since the OSC functions do not include a
communicator argument and thus would not recognize a revoked
communicator. After the revoke, \emph{all} processes know of the failure,
which is not strictly necessary for our application.

Afterwards, the processes could avoid communication with the
failed process and the affected window blocks. From the
application logics, the recovery could be performed on the intact
processes with the help of the saved data and their replicas (which
would be copied and redistributed to replace lost data).

Unfortunately, this is prevented by the following problem: Assume,
process~X locks a window on a surviving
process~P1, and then~X dies. Naturally, no other process knows about the locking. Consequently, some
other process~P2 may try to lock the same window afterwards. This
\verb|MPI_Win_lock| call blocks until the window is unlocked, which will
never happen because~X has died. Even if the program was
able to recognize the situation, it could not get out of it, since the lock
and unlock functions must be called by the same process, but~X is dead.  For
cases like that, it would be helpful to have an enforced unlocking
function, or at least a lock function with a timeout parameter.

For now, we chose a different approach: Each process that learns about
the failure, issues an \verb|MPI_Comm_Shrink| call. At its return, the
process has entered transaction mode.  To ensure a consistent view of
the set of lost processes, the processes afterwards invoke
\verb|MPI_Comm_agree| and determine the lost ranks with the help of
the process groups of the original and shrinked communicators (using
function \verb|MPI_group_translate_ranks|).

Thereafter, \verb|MPI_Win_free| is called on the damaged window.
Luckily, although rather surprisingly, this call works in our experience, despite any held locks and
lost processes. Thus, afterwards, the processes of the new
communicator can collectively create a new window. Since each process
can access its local window block with normal C/C++ operations, it can easily copy the window data;
or, even simpler, it can specify the same memory
block for the new window avoiding the need for copying. Similarly,
memory blocks for the lost data copies can be added by collective operations. Their distribution
should again follow some fixed pattern, but the
particular pattern is not relevant for this paper.

Summarizing the above, to repair the window, we had to repair the
communicator first, which feels like a detour. A more direct way to
shrink a window would have simplified programming.

For the realization of transactions, we adopted a variant of the
two-phase commit (2PC) protocol~\cite{15ausRai}
from~\cite{saraThesis}. Briefly stated, each 2PC transaction is
supervised by a \emph{transaction coordinator}. The coordinator first requests
the locking and checking of all required data copies (prepare
phase), and afterwards the actual changes (commit
phase).

For simplicity, our store implementation uses a single
transaction coordinator, namely process~0. All processes send their
transactions to this coordinator, which executes them in
sequence. Also for simplicity, the coordinator exclusively locks all
memory blocks at entrance to transaction mode. While this design
is not optimal regarding perfomance, it still allows uninvolved worker processes
of the application to continue, as long as they need not access the
store.  To handle a possible loss of the above (master) transaction
coordinator, backup coordinators are designated. The master
coordinator keeps them up to date about transaction progress, so that
they can take over if needed.

The transaction mode is left by a collective operation that must be
invoked by the application after recovery.

\subsection{Resilient Store with Localized Recovery}

Our application would actually be able to perform a localized recovery (see
Sect.~\ref{sect:reqs}), but the above usage of collective operations
prohibits that. These operations were introduced because a
window becomes unusable after having been locked by a failed process.

In the resilient store design described in this section, the MPI
locks, which cause the above problems, are replaced by
application-level locks realized with \verb|MPI_Compare_and_swap|
(CAS): To lock a window, a process atomically changes some lock
variable from~-1 to its own number; and to unlock it, the process
changes it back from its own number to~-1. If the variable does not
contain~-1 initially, the process repeats the CAS call. Since CAS
returns the current variable value, the calling process can compare
this value with the ids of failed processes (or ping the owner), and
unlock the memory by itself if needed.

Still, other programming problems remain. Most importantly, memory
blocks at the failed processes must be restored at surviving processes, which requires new
memory allocations. Creating new windows is
impossible without a collective call, but a collective call can not be
issued in the damaged old communicator. Enlarging the existing blocks
is not possible either, as has been discussed above.

On the positive side, consistent use of CAS in all resilient store
access operations eliminates the need for a distinction into normal
and transaction modes. Transactions can be performed at all times,
when they use the same locks as the standard operations. Still, use of
locking, even through CAS, takes more time than simple
\verb|MPI_Put/Win_flush| calls.

While a resilient store along these lines is certainly possible, its
implementation seems demanding, notably because the identities of
failed ranks must be kept record of. Moreover, such a store design
does not follow the OSC design philosophy of partitioning the program
execution into independent epochs. Instead, it uses a single epoch,
invoking \verb|MPI_Win_lock_all(MPI_MODE_NOCHECK, ...)| once after
window creation, and \verb|MPI_Win_Unlock| right before
\verb|MPI_Win_free|. For such a design, the locking/unlocking could
better have been coupled with window creation/destruction.

\subsection{Summary}
Altogether, we missed the following RMA/ULFM functionalities:
\begin{itemize}
\item blocking variant of \verb|MPI_Put|,
\item window creation/deletion functions with implicit epoch definition,
\item signaling mechanism to allow one process to interrupt another,
\item failure propagation to a subset of processes,
\item enforced unlocking function or locking function with timeout,
\item function for shrinking a window without shrinking the communicator,
\item error message or warning issued when \verb|MPI_Win_free| is called on a locked window, and
\item query function for revocation state of a communicator that works
  without prior communication.
\end{itemize}
 
\section{Experiences with MPI RMA and ULFM Usage}\label{sect:experiences}

Despite the warning depicted at the end of
Section~\ref{sect:background}, we experimented with ULFM usage in OSC
programs. Experiments were run with different Open MPI versions,
such as version 5.0.7, on different machines and with different
\verb|mpirun| arguments (but always including
\verb|--withft ulfm|). Observations were partially inconsistent, but
essentially confirmed that Open MPI is not yet ready for OSC+ULFM
usage.

Two examples of unexpected behaviors are outlined
in Listings~\ref{lst:lst2} and \ref{lst:lst1}. The program in
Listing~\ref{lst:lst2} works fine when the \verb|sleep| call is omitted,
but with \verb|sleep|, it issues warnings about OSC usage and does
not terminate. Thereby \emph{no} process was killed.

The
program in Listing~\ref{lst:lst1} reports the process
abort, but then continues producing the regular output. Thus,
the window block of the failed process seems to be further used despite the process's
failure. As a third example, killing a PRTE demon at one node of a
multi-node execution (to simulate node failure) crashed the overall
execution.

{\small
\begin{lstlisting}[caption={Program does not terminate, despite no process aborts},label=lst:lst2]
    ...
    MPI_Comm_set_errhandler(comm, MPI_ERRORS_ARE_FATAL);
    MPI_Win_set_errhandler(win, MPI_ERRORS_ARE_FATAL);

    MPI_Win_lock(MPI_LOCK_EXCLUSIVE, 1, 0, win);
    constexpr int i = 1;
    MPI_Put(&i, 1, MPI_INT, 1, 0, 1, MPI_INT, win);
    MPI_Win_unlock(1, win);
    MPI_Barrier(comm);
    sleep(10);
    MPI_Win_free(&win);
    MPI_Finalize();
\end{lstlisting}
}

{\small
\begin{lstlisting}[caption={Program generates regular output, despite process abort},label=lst:lst1]
    ...
    MPI_Win_set_errhandler(win, MPI_ERRORS_ARE_FATAL);

    if (rank == 1) {
        MPI_Win_lock(MPI_LOCK_EXCLUSIVE, 1, 0, win);
        constexpr int buf = 123;
        MPI_Put(&buf, 1, MPI_INT, 1, 0, 1, MPI_INT, win) 
        MPI_Win_unlock(1, win);
        abort();
    } else {
        sleep(5); // wait for process to terminate
        MPI_Win_lock(MPI_LOCK_EXCLUSIVE, 1, 0, win);
        int buf = -1;
        MPI_Get(&buf, 1, MPI_INT, 1, 0, 1, MPI_INT, win);
        MPI_Win_flush(1, win);
        printf("Data flushed after get: %d\n", buf);
        buf = 456;
        MPI_Put(&buf, 1, MPI_INT, 1, 0, 1, MPI_INT, win);
        MPI_Win_flush(1, win) != MPI_SUCCESS);
        buf = -1;
        MPI_Get(&buf, 1, MPI_INT, 1, 0, 1, MPI_INT, win);
        MPI_Win_flush(1, win);
        printf("Data flushed after get: %d\n", buf);
        MPI_Win_unlock(1, win);
    }
    MPI_Win_free(&win);
    MPI_Finalize();

    // output (abbreviated):
    // ...
    // Signal: Aborted (6)
    // ...
    // Data flushed after get: 123
    // Data flushed after get: 456
    // Data flushed after get: 456
    // <terminates regularly>
\end{lstlisting}
}

Moreover, we tested all OSC functions under \verb|MPI_ERRORS_RETURN|
and enforced process aborts. The only function that reported
\verb|MPI_ERR_PROC_FAILED| was \verb|MPI_Win_fence|. All other
functions exhibited non-terminating behavior when the communication
partner was dead. From a practical point of view, such non-terminating
behavior is even worse than a program abort, and timeouts would have
helped here.

\section{Workaround and Outline of Implemented Store}\label{sect:workaround}

From Section~\ref{sect:experiences}, we could not rely on ULFM error
reporting from OSC functions in the implementation of our store.
Therefore, we adopted a workaround, preceding each OSC call with a
ping. Unfortunately, ULFM does not explicitly support such a
ping. Therefore, we experimented with different two-sided functions
that would not disturb the progress of our application, such as
\verb|MPI_Probe| and \verb|MPI_Test|, but none returned an error code.
We could have employed a ping at a lower layer, such as in UCX, but we
strived for a portable solution.

Therefore, we decided to realize the ping in a helper thread,
employing one background thread per process.  This thread communicates
with the main thread through OpenMP~\cite{openmp}, and we
initialize the MPI system with \verb|MPI_THREAD_MULTIPLE|. Prior to
each OSC call, the main thread issues an \verb|MPI_Send| to the
target, where it matches an \verb|MPI_Iprobe(...ANY_SOURCE...)| in the
helper thread.  This \verb|MPI_Iprobe| is preventively posed in a
loop, and followed by \verb|MPI_Recv()| upon message arrival.
\verb|MPI_Send| is a two-sided operation, and correctly returns an error
code in the event of a failure. If a failure is reported, the planned OSC
function call is omitted and the recovery procedure from
Section~\ref{sect:design} initiated instead.

Note that there is still a certain risk that the target process aborts
after the \verb|MPI_Send|, but before the OSC function. Then, the OSC call, and
thus the overall program execution, does not terminate. It does not
seem possible to fully avoid this risk. A program abort
would certainly be preferable to non-termination, and therefore we tried usage of two
communicators: one with \verb|MPI_ERRORS_RETURN| for the two-sided
pings, and another with \verb|MPI_ERRORS_ARE FATAL| for the OSC
functions. Unfortunately, the latter did not cause program abort either,
similar to the example in Listing~\ref{lst:lst1}.

The above workaround impairs the performance, due to the additional
calls and threads. However, it has the nice side effect
of enabling window enlargement: If a source process discovers a window
block of insufficient size, it can now send a message into the posted
receive of the helper thread at the target, and the target can
react. Similarly, the helper thread can assist in updating the backup
transaction coordinators in transaction mode.

We performed some initial performance measurements on Infiniband-connected nodes of the Goethe-NHR cluster
at the University of Frankfurt~\cite{ClusterGoethe}.  Each node
is equipped with two 20-core Intel Xeon Skylake Gold 6148
CPUs and 192~GB of main memory.  We assigned $40$~worker processes to
each node and used Open MPI version~5.0.7.

\begin{figure}[ht]
	\centering
	\begin{minipage}[t]{0.455\linewidth}
		\centering
		\includegraphics[width=\linewidth]{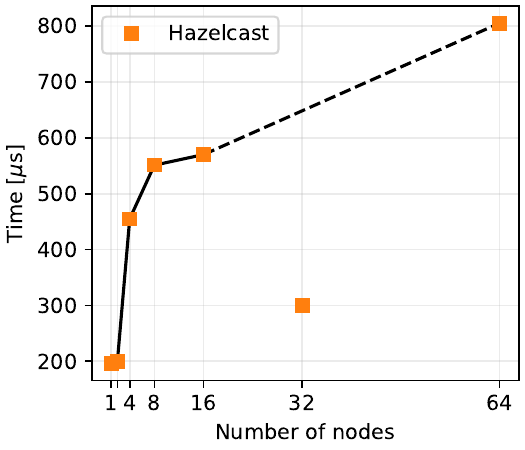}
                 \caption{store\_put latencies of Hazelcast, with 6 replicas and 40 processes per node}
		\label{fig:hazelcast-measurements}
	\end{minipage}
	\hfill
	\begin{minipage}[t]{0.455\linewidth}
		\centering
		\includegraphics[width=\linewidth]{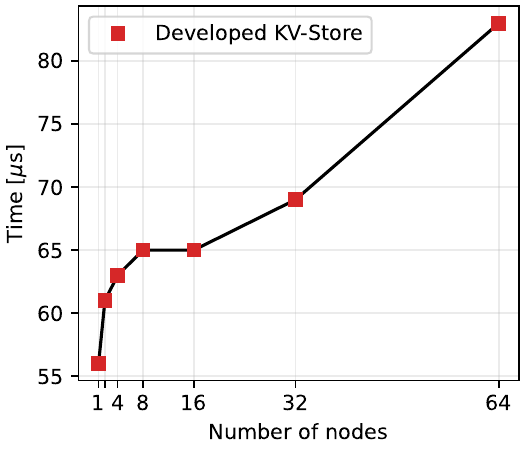}
                \caption{store\_put latencies of our store, with 6 replicas and 40 processes per node}
		\label{fig:kv-store-measurements}
	\end{minipage}
\end{figure}

Results are depicted in Figure~\ref{fig:kv-store-measurements}, and
compared to corresponding results obtained with the Hazelcast IMap in
Figure~\ref{fig:hazelcast-measurements}. The runs
have been carried out with slightly different benchmarks (but on the same
machine), and are thus not fully comparable. Nevertheless, the shown
performance improvement by a factor of almost~10 (see scales of the
figures) can for the most part be attributed to the specialized design
of our new store.

The above experiments were conducted in failure-free mode, but with \verb|with-ft ulfm|.
Furthermore we run a few initial correctness tests after process
aborts, but further tests and a performance evaluation of the recovery process
are still open.

\section{Related Work}\label{sect:related}

Prior to its integration into Open MPI, ULFM was provided as an
independent software package~\cite{altesULFM}. The system has been
widely used, for instance for the realization of Algorithm-Based Fault
Tolerance (e.g.~\cite{abftBsp2,abftBsp1}). Most ULFM applications
perform a non-shrinking global recovery~\cite{implicit22}, but
localized recovery strategies have been followed as well
(e.g.~\cite{ulfmFGCS,aurelienAbstract}).  An important application of
ULFM has been the implementation of Resilient
X10~\cite{saraULFM}. Therein collective and two-sided communication is
protected, and \verb|MPI_Iprobe| is used in a similar way as in this
paper.

Alternatives to ULFM include Fenix~\cite{fenix} and
Reinit++~\cite{reinit}. Reinit++ only allows global-restart
recovery. A variant of Fenix allows localized recovery, but failed
processes need to be replaced by spare
processes~\cite{ftx1}. Fenix includes a data recovery
component, and recent work sketches the combination of Fenix and ULFM for
``pseudo-local'' checkpointing and recovery~\cite{aurelienAbstract}.

Several shortcomings of RMA have been discussed
in~\cite{quoVadis}. They include problems with window enlargement, but
other problems than discussed in this paper.  A notification mechanism
for MPI enables the attachment of callbacks to requests and might help
with our window enlargement problem~\cite{mpiContinue}.

Other resilient store options include third-party stores such as the
Hazelcast IMap~\cite{hazelcast}, Redis~\cite{redis}, etcd~\cite{etcd}
and GridGain~\cite{gridGain}. Finally, instead of a ``real'' store,
vulnerable data can be saved on MPI process~0, and the program be
aborted if this process dies~\cite{unserFGCS}.

\section{Conclusions}\label{sect.conclusions}

This paper has considered the realization of an
application-specific resilient store with MPI OSC and ULFM. We have
discussed design options and OSC/ULFM support.  Thereby we listed
functionalities that we missed, such as window enlargement and the
unlocking of a window after abort of the lock owner. Moreover, we
reported some experiences with the current ULFM implementation. Since
OSC functions do not yet return error codes, we adopted a workaround
using two-sided communication and a helper thread. Our implementation is functional, but the
additional two-sided function calls of our workaround cause overheads. Moreover, the store may still hang, namely when a process aborts between the
two-sided and one-sided calls.

ULFM support for OSC would be highly desirable. Regarding the resilient
store implementation itself, future work includes further tests and performance
measurements, as well as an implementation of the localized recovery
variant.

\section*{Acknowledgements} This research was funded by the Deutsche Forschungsgemeinschaft (DFG, German
Research Foundation) under project number~512078735.
The authors gratefully acknowledge the computing time provided to them on the
Goethe-HLR cluster at the Frankfurt Center for Scientific Computing.
Thanks to Daniel Szumski for conducting the Hazelcast measurements.

\bibliographystyle{splncs04}
\bibliography{bibo}
\end{document}